\pdfoutput=1

\documentclass[11pt,twoside,a4paper,cmspaper,final,collab]{cms-tdr}
\def\svnVersion{261923}\def\cmsCernNo{2013-037}\def\cmsCernDate{\today}\def\cmsMessage{Published in Physical Review Letters as \href{http://dx.doi.org/10.1103/PhysRevLett.113.132301}{\doi{10.1103/PhysRevLett.113.132301.}}}

\begin{document}\cmsNoteHeader{HIN-12-003}

\hyphenation{had-ron-i-za-tion}
\hyphenation{cal-or-i-me-ter}
\hyphenation{de-vices}

\RCS$Revision: 136722 $
\RCS$HeadURL: svn+ssh://svn.cern.ch/reps/tdr2/notes/HIN-12-003/trunk/HIN-12-003.tex $
\RCS$Id: HIN-12-003.tex 136722 2012-07-12 19:14:34Z ssalur $
\newlength\cmsFigWidth
\ifthenelse{\boolean{cms@external}}{\setlength\cmsFigWidth{0.85\columnwidth}}{\setlength\cmsFigWidth{0.4\textwidth}}
\ifthenelse{\boolean{cms@external}}{\providecommand{\cmsLeft}{top}}{\providecommand{\cmsLeft}{left}}
\ifthenelse{\boolean{cms@external}}{\providecommand{\cmsRight}{bottom}}{\providecommand{\cmsRight}{right}}
\providecommand{\raa}{\ensuremath{R_\mathrm{AA}}\xspace}
\providecommand{\taa}{\ensuremath{T_\mathrm{AA}}\xspace}
\cmsNoteHeader{HIN-12-003} 

\title{\texorpdfstring{Evidence of \cPqb-jet quenching in PbPb collisions at $\sqrt{s_{NN}} = 2.76$\TeV}{Evidence of b-jet quenching in PbPb collisions at sqrt(s[NN]) = 2.76 TeV}}

\date{\today}

\abstract{
The production of jets associated to bottom quarks is measured for the first time in PbPb collisions at a center-of-mass energy of 2.76\TeV per nucleon pair. Jet spectra are reported in the transverse momentum (\pt) range of 80--250\GeVc, and within pseudorapidity $\abs{\eta} < 2$. The nuclear modification factor ($\raa$) calculated from these spectra shows a strong suppression in the \cPqb-jet yield in PbPb collisions relative to the yield observed in $\Pp\Pp$ collisions at the same energy. The suppression persists to the largest values of \pt studied, and  is centrality dependent.  The $\raa$ is about 0.4 in the most central events, similar to previous observations for inclusive jets. This implies that jet quenching does not have a strong dependence on parton mass and flavor in the jet \pt range studied.
}

\hypersetup{%
pdfauthor={CMS Collaboration},
pdftitle={Evidence of b-jet quenching in PbPb collisions at sqrt(s[NN]) = 2.76 TeV},
pdfsubject={CMS},
pdfkeywords={CMS, physics, heavy ions, jet quenching, b tagging, PbPb}}

\maketitle

By colliding heavy nuclei at the Large Hadron Collider (LHC), one expects to reach sufficiently large energy densities to form a strongly coupled quark-gluon plasma (QGP), a state which is characterized by effective deconfinement of the quarks and gluons~\cite{Shuryak:1977ut,Collins:1974ky,Shuryak:2004cy}. Hard-scattered partons are expected to suffer energy loss as they traverse the QGP via elastic and inelastic interactions~\cite{Bjorken:1982tu,Baier:2000mf}. This is commonly thought to be the mechanism responsible for the observed suppression of high transverse momentum (\pt) hadrons and jets, or ``jet quenching'', in nuclear collisions~\cite{Adcox:2004mh,Adams:2005dq,Back:2004je,Arsene:2004fa, Adams:2003im,Adams:2005ph, Adler:2003qi, Adare:2008ae,Chatrchyan:2012nia,Collaboration:2010bu,Chatrchyan:2011sx}.
Measurements of parton energy loss are expected to reveal the fundamental thermodynamic and transport properties of this phase of matter (see Refs.~\cite{CasalderreySolana:2007zz,dEnterria:2009am} for recent reviews).

The quenching of jets in heavy-ion collisions is expected to depend upon the flavor of the fragmenting parton.  Energy loss via gluon bremsstrahlung, which is thought to be the dominant mechanism for light partons, should be larger for gluon jets than quark jets, due to the larger color factor for gluon emission from the former.  The mass of the leading parton may also play a role.  Collisional energy loss could be an important effect for massive quarks and has been invoked to describe the nuclear modification factors for leptons from heavy-flavor decays at low \pt~\cite{Wicks:2005gt, Adare:2006nq, Adare:2010de}.  In this regime, radiative energy loss from heavy quarks may be suppressed due to coherence effects~\cite{Dokshitzer:2001zm, Armesto:2003jh}, although the relevance of such effects in finite-size systems is a subject of debate~\cite{Aurenche:2009dj}.  The strongly coupled nature of the QGP may also introduce mass effects, according to a description of jet quenching based on the AdS/CFT correspondence~\cite{Gubser:2006bz, Horowitz:2007su}.  Consequently, measurements of the flavor and mass dependence of jet quenching are essential to obtain a sound theoretical description of this phenomenon.

Measurements of hadrons containing $\cPqb$ quarks are expected to be sensitive to the details of $\cPqb$-quark energy loss.  Recent data on single-particle production of $\PB$ mesons (via non-prompt $\JPsi$)~\cite{Chatrchyan:2012np} show a smaller suppression compared to $\PD$ mesons~\cite{ALICE:2012ab} and non-identified charged particles~\cite{hin10005,Abelev:2012hxa}.  Experimentally, the jet associated to a $\cPqb$ hadron is commonly referred to a ``$\cPqb$ jet'', although the $\cPqb$ quark is not guaranteed to be the leading parton of the jet.  In relation to $\PB$ mesons, $\cPqb$ jets provide a complementary approach to study \cPqb-quark energy loss, albeit typically in a different range of \pt. Through comparisons with the existing measurements of inclusive jet production~\cite{Aad:2012vca}, $\cPqb$-jet measurements can be used to study the flavor dependence of jet quenching, which in turn provides insight on the dynamics of parton energy loss.

The Compact Muon Solenoid (CMS) detector has excellent capabilities to perform \cPqb-jet identification ($\cPqb$ tagging) measurements as demonstrated in Ref.~\cite{Chatrchyan:2012jua}. Measurements of the \cPqb-jet cross section~\cite{Chatrchyan:2012dk} and \cPqb-jet angular correlations~\cite{Khachatryan:2011wq} have been performed in $\Pp\Pp$ collisions at 7\TeV.  This Letter presents the first measurements of \cPqb-jet production in heavy-ion collisions using a dataset corresponding to an integrated luminosity of 150 $\mubinv$ of PbPb collisions at $\sqrt{s_{_{NN}}}=2.76$ TeV delivered by the LHC in 2011. The comparison measurements are performed with a dataset consisting of $\Pp\Pp$ data recorded in 2013 and corresponding to an integrated luminosity of 5.3\pbinv at $\sqrt{s}=2.76$\TeV.

The central feature of the CMS apparatus is a superconducting solenoid providing a magnetic field of 3.8\unit{T}. Charged particle trajectories are measured with the silicon tracker, which provides an impact parameter resolution of ${\sim}15\mum$ and a \pt resolution of ${\sim}1.5$\% for 100\GeVc particles. A PbWO$_{4}$ crystal electromagnetic calorimeter (ECAL) and a brass/scintillator hadron calorimeter (HCAL) surround the tracking volume.  The forward regions ($2.9 < \abs{\eta} < 5.2$, where $\eta = - \ln[\tan (\theta/2)]$ and $\theta$ is the polar angle measured with respect to the counterclockwise beam direction) are instrumented with iron/quartz-fiber hadron forward calorimeters (HF). Collision centrality, defined as a percentile of the total inelastic nucleus-nucleus cross section, is calculated using the sum of the HF transverse energy~\cite{Chatrchyan:2011pb}. A set of scintillator tiles, used for triggering and beam-halo rejection, is mounted on the inner side of the HF calorimeters. A more detailed description of the CMS detector can be found in Ref.~\cite{cmsexp}.

Jets are reconstructed from particle candidates obtained from a particle-flow algorithm~\cite{CMS-PAS-PFT-09-001}.
This algorithm improves the resolution of jets, while reducing the parton flavor dependence of the detector response as compared to a purely calorimetric measurement.
The anti-\kt clustering algorithm~\cite{Cacciari:2008gp} is used, with a distance parameter of $R=0.3$. Details of the jet reconstruction, resolution and energy corrections may be found in Refs.~\cite{Chatrchyan:2011sx,Chatrchyan:2012nia,Chatrchyan:2012gt}. The underlying background of bulk particle production in PbPb collisions is subtracted using the same method described in Ref.~\cite{Kodolova:2007hd}. Jet \pt\ resolution effects are unfolded using an iterative method ~\cite{UNFOLD}, as implemented in the \textsc{RooUnfold} package~\cite{ROOUNFOLD}.

The Monte Carlo simulations are performed using {\PYTHIA} 6.422~\cite{bib_pythia} with tune Z2~\cite{Field:2010bc}.
A parton flavor is assigned to reconstructed jets by matching them in  $\Delta R = \sqrt{\smash[b]{{\Delta\eta}^{2} + {\Delta\phi}^{2}}}$ to generator-level partons ($\phi$ is the azimuthal angle measured in radians in the plane transverse to the beams).  If a bottom quark is found within $\Delta R < 0.3$ then the jet is considered to be a $\cPqb$ jet, irrespective of any other partons in the cone. This definition includes b quarks from gluon splitting ($\Pg \rightarrow \bbbar$), even if the splitting occurs late in the parton shower (\ie, at low virtuality), consistent with the theoretical treatment of heavy-flavor production in Ref.~\cite{Frixione:1996nh,Cacciari:2012ny}.  We note that $\cPqb$ jets from gluon splitting comprise about 30--35\% of the total \cPqb-jet cross section according to \PYTHIA simulations, although measurements of $\cPqb-\bar{\cPqb}$ angular correlations at 7 TeV indicate that contribution is somewhat larger~\cite{Khachatryan:2011wq}.  Such jets are expected to interact differently with the QGP than those from primary b quarks~\cite{Huang:2013vaa}. To compare with PbPb data, \PYTHIA events are embedded into PbPb events produced by the \textsc{hydjet} generator (version 1.8) \cite{Lokhtin:2005px}, which is tuned to reproduce event properties, such as charged-hadron multiplicity, \pt\ spectra, and elliptic flow.  The rate of bottom quark production per nucleon-nucleon interaction in \textsc{hydjet} was found to be consistent with theoretical calculations for pp collisions based on Ref.~\cite{Cacciari:2012ny}.

\begin{figure*}[htb]
\centering
\includegraphics[width=0.45\textwidth]{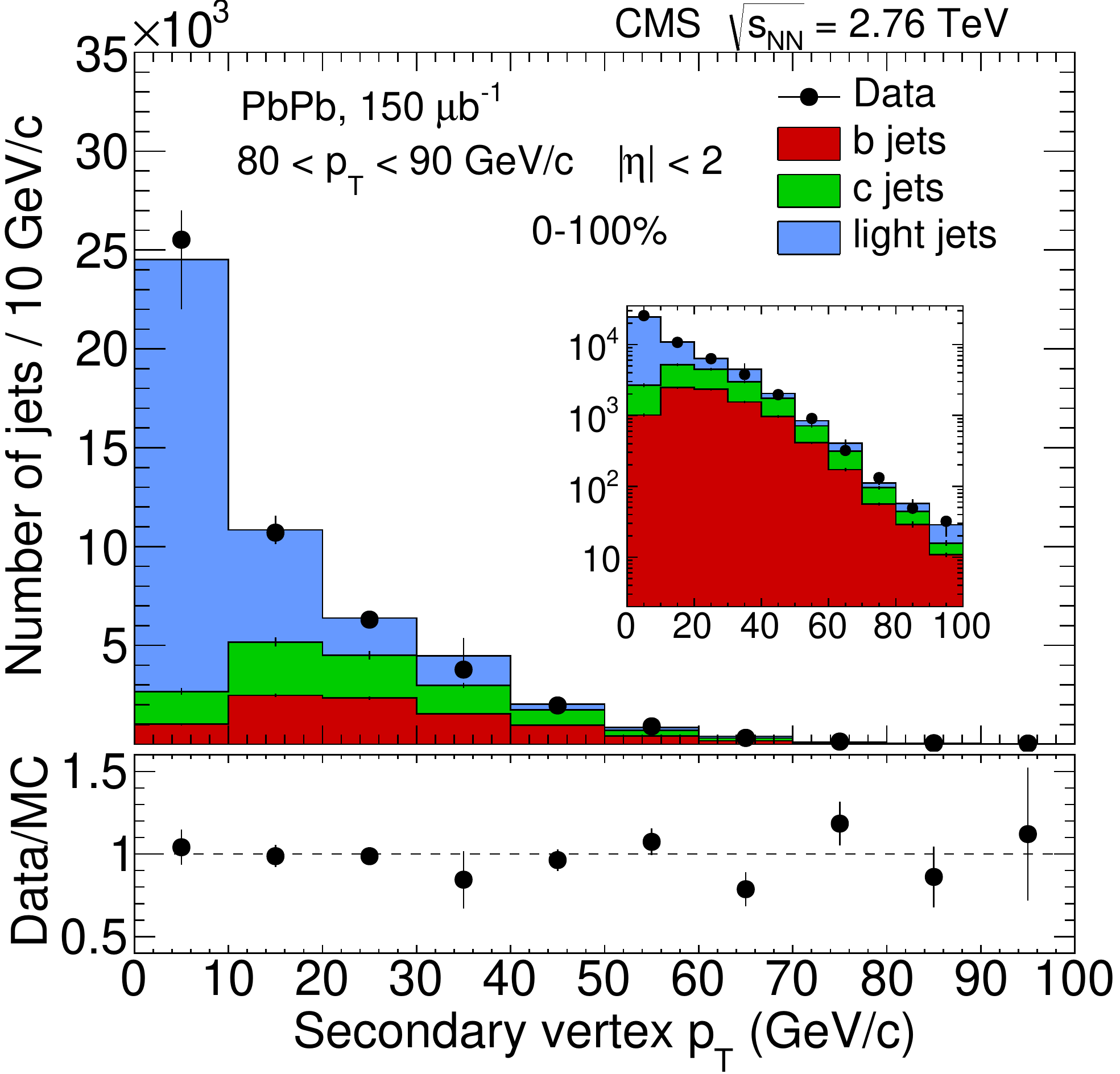}\hspace*{0.05\textwidth}
\includegraphics[width=0.45\textwidth]{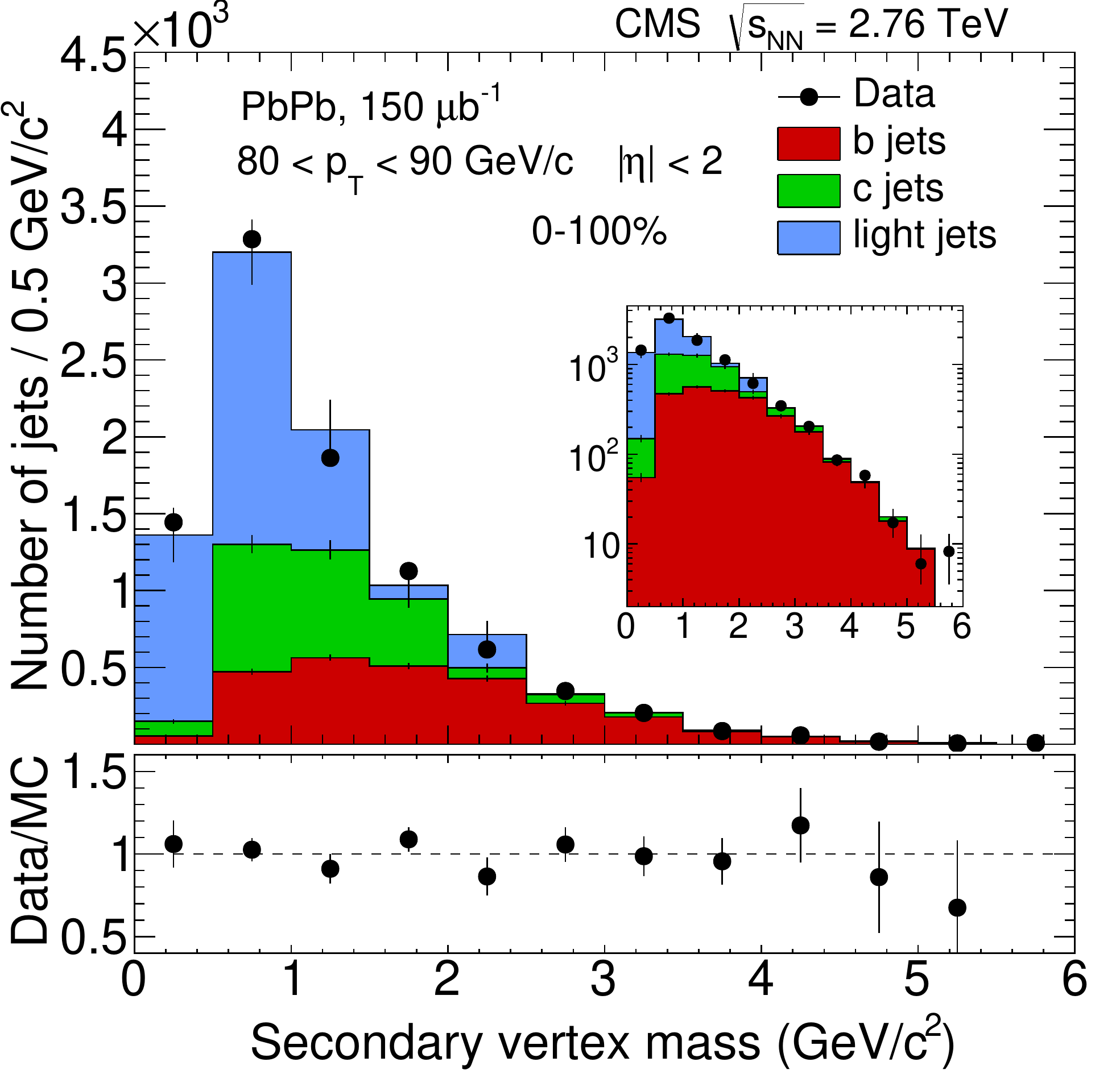}
\caption{\label{fig:fitPbPb_SVmass}
Left: Comparison of SV \pt distribution between data and simulation for the same jet and event selections.  The simulation is normalized to the data. Right: Template fit to the SV invariant mass distribution in centrality integrated (0--100\%) PbPb collisions for jets of $80 < \pt < 90$\GeVc.  Insets show the same comparisons with the y-axis in log scale.
}
\end{figure*}

Identification of $\cPqb$ jets is based on kinematic variables related to the relatively long lifetime and large mass of b hadrons. Charged tracks of \pt $> 1$\GeVc within $R < 0.3$ from the jet axis are used to reconstruct secondary vertices (SV) from $\cPqb$ hadrons and/or subsequent \cPqc-hadron decays from the $\cPqb\to\cPqc$ cascade, using an adaptive vertex fit~\cite{Fruhwirth:2007hz}. The contribution of $\cPqb$ jets is enhanced by requiring that SVs are far enough from the primary vertex, using a selection on the significance of the three-dimensional flight distance. This selection is chosen to give a misidentification rate of roughly 1\% for light jets and 10\% on charm-quark jets ($\cPqc$ jets), based on simulation. The corresponding \cPqb-tagging efficiency is about 65\% for $\Pp\Pp$ and 45\% for PbPb collisions.  The compatibility of the simulation with data was verified by comparing basic distributions such as the $\chi^{2}$ of the SV fit, the number of tracks per SV, and the number of SVs per jet.  Figure~\ref{fig:fitPbPb_SVmass} (left) shows an example comparison of the SV \pt distribution.  The shape of the distribution is well described over the full \pt range.

The SV invariant mass is calculated from the constituent tracks. An example SV mass distribution, for jets with $80 < \pt < 90\GeVc$, is shown in Fig.~\ref{fig:fitPbPb_SVmass} (right).  For each jet \pt\ bin, the \cPqb-jet purity ($f_\cPqb$), \ie, the ratio of the number of $\cPqb$ jets to that of inclusive jets in the tagged sample, is extracted by means of a template fit. The shapes of the light-quark, $\cPqc$ and $\cPqb$ contributions are determined from simulation, while their normalizations are allowed to float. After tagging, the three contributions are of comparable magnitude, as shown in the figure, but the \cPqb-quark contribution dominates above the \cPqc-quark mass threshold near 2\GeVcc, which allows for an accurate determination of the \cPqb-jet contribution.  The quality of the SV mass fits was found to be good, with values of $\chi^{2}$ per degree of freedom typically in the range of 1--2.  The proportion of tagged jets for which the SV corresponds to a $\cPqb$ hadron from a different nucleon-nucleon interaction than the one that produced the jet was estimated from simulation to be 2\% for the 20\% most central PbPb collisions.

For the systematic studies described below, an alternative b-tagging strategy is employed, which uses the jet probability (JP) algorithm~\cite{Chatrchyan:2012jua}. In contrast to direct reconstruction of SVs, the JP tagger is based on an estimate of the compatibility of tracks with the primary vertex, using their three-dimensional impact parameter significance. A probability density for this compatibility is obtained directly from data using tracks with negative impact parameter, which are unlikely to come from heavy-flavor decays. The impact parameter (IP) is defined to have the same sign as the scalar product of the vector pointing from the primary vertex to the point of closest approach with the jet direction. Tracks originating from the decay of particles traveling along the jet axis will tend to have positive IP values.

Using the \cPqb-jet purity ($f_\cPqb$) derived from the template fit, the \cPqb-jet yield in a given \pt\ bin is obtained as $N_\cPqb = N f_\cPqb / \epsilon$, where $N$ is the number of all \cPqb-tagged jets and $\epsilon$ is the \cPqb-tagging efficiency.  The efficiency $\epsilon$ is determined from simulation and cross-checked using the so-called reference lifetime tagger method~\cite{Chatrchyan:2012jua}, which uses the JP tagger to determine the efficiency of the SV tagger directly from data, taking advantage of the calibration of the primary vertex compatibility used in this tagger which is obtained from data. The simulation reproduces the estimate of $\epsilon$ from data to within 5\%.

\begin{figure}[htb]
\begin{center}
\includegraphics[width=0.45\textwidth]{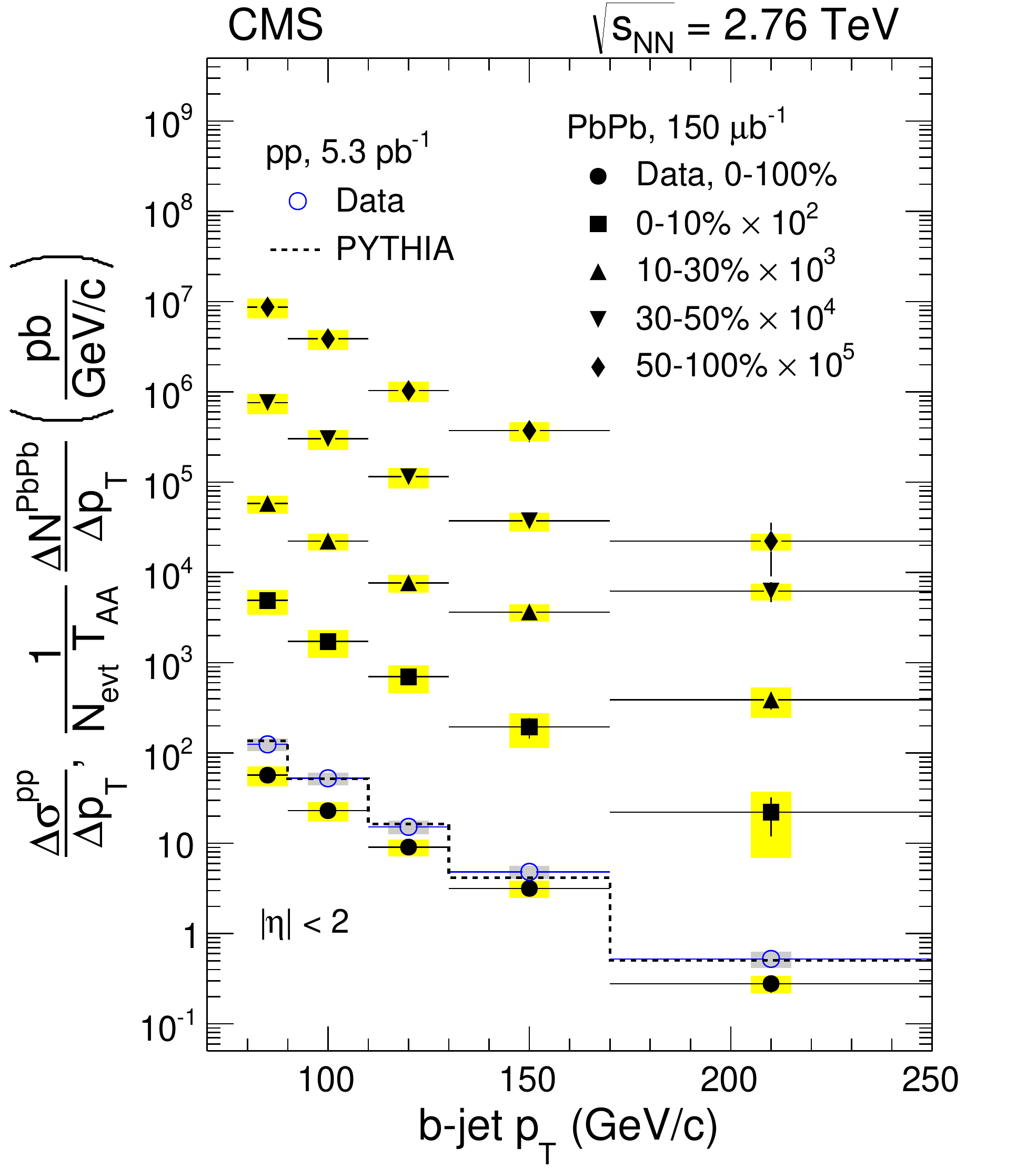}
\caption{\label{fig:spectra}
The \cPqb-jet yield as a function of \pt\ is shown for various centrality classes of PbPb collisions as indicated in the legend. The yields are scaled by the equivalent number of minimum bias events sampled and by  \taa. The spectra are also scaled by powers of 10 for visibility.  The \cPqb-jet cross section in $\Pp\Pp$ collisions is also shown, and compared to \PYTHIA. Vertical and horizontal bars represent statistical uncertainties and bin widths, respectively, while filled boxes represent systematics uncertainties.
}
\end{center}
\end{figure}

The unfolded \cPqb-jet \pt\ spectra in PbPb collisions are shown in Fig.~\ref{fig:spectra} for several centrality selections. The PbPb data are divided by \taa, computed from a Glauber model (for a review, see Ref.~~\cite{Miller:2007ri}), to scale to the expectation for $\Pp\Pp$ collisions in the absence of nuclear effects.  The value of \taa is the number of nucleon-nucleon (NN) collisions divided by the total inelastic NN cross section and may be interpreted as the NN equivalent luminosity per PbPb collision.  Also shown is the measured \cPqb-jet cross section in $\Pp\Pp$ collisions.  The cross section is compared to {\PYTHIA} simulations, which agree well with the data, as is the case at $\sqrt{s}=7$\TeV for the \pt range covered by the present study~\cite{Chatrchyan:2012dk}.

The systematic uncertainties fall into two general categories: $\cPqb$ tagging and jet reconstruction. The  \cPqb-tagging uncertainty on \cPqb-jet yields varies from about 12 to 18\%,  depending on jet \pt\ and collision system. The uncertainty is evaluated via the following systematic variations of the tagging procedure, which influence the extracted  \cPqb-tagging purity and efficiency values:

\begin{itemize}
 \setlength{\itemsep}{1pt}
  \setlength{\parskip}{0pt}
  \setlength{\parsep}{0pt}
\item varying the SV flight distance selection such that $\epsilon$ differs by about 10\%,
\item using $\epsilon$ from the reference lifetime tagger method~\cite{Chatrchyan:2012jua}, rather than from simulation,
\item fixing the $\cPqc$ jet to light-quark jet normalization, rather than allowing them to float independently in the template fits,
\item using a non-\cPqb-jet template produced from jets with small JP in data,

\item varying the gluon-splitting contribution in the \cPqb-jet and \cPqc-jet templates by 50\%.

\end{itemize}

The uncertainty on the spectra due to the jet reconstruction is 10--12\% for $\Pp\Pp$ and 15--17\% for PbPb, and is comprised of the following sources:
\begin{itemize}
 \setlength{\itemsep}{1pt}
  \setlength{\parskip}{0pt}
  \setlength{\parsep}{0pt}
\item a 10\% uncertainty in the jet energy resolution~\cite{CMS-PAS-JME-10-010},
\item a 2\% uncertainty in the jet energy scale (JES)~\cite{CMS-PAS-JME-10-010},
\item an additional, centrality-dependent, 1--2\% uncertainty in the JES in PbPb collisions due to the underlying event, evaluated from random-cone and embedding studies,
\item an uncertainty in the unfolding procedure evaluated by varying the number of iterations and the presumed prior spectrum.
\end{itemize}

The $\Pp\Pp$ luminosity has an uncertainty of 3.6\%, while the uncertainty in  \taa  varies from about 4\% for centrality 0--10\% to 15\% for 50--100\%~\cite{Chatrchyan:2011sx}.

\begin{figure}[htb]
\begin{center}
\includegraphics[width=0.48\textwidth]{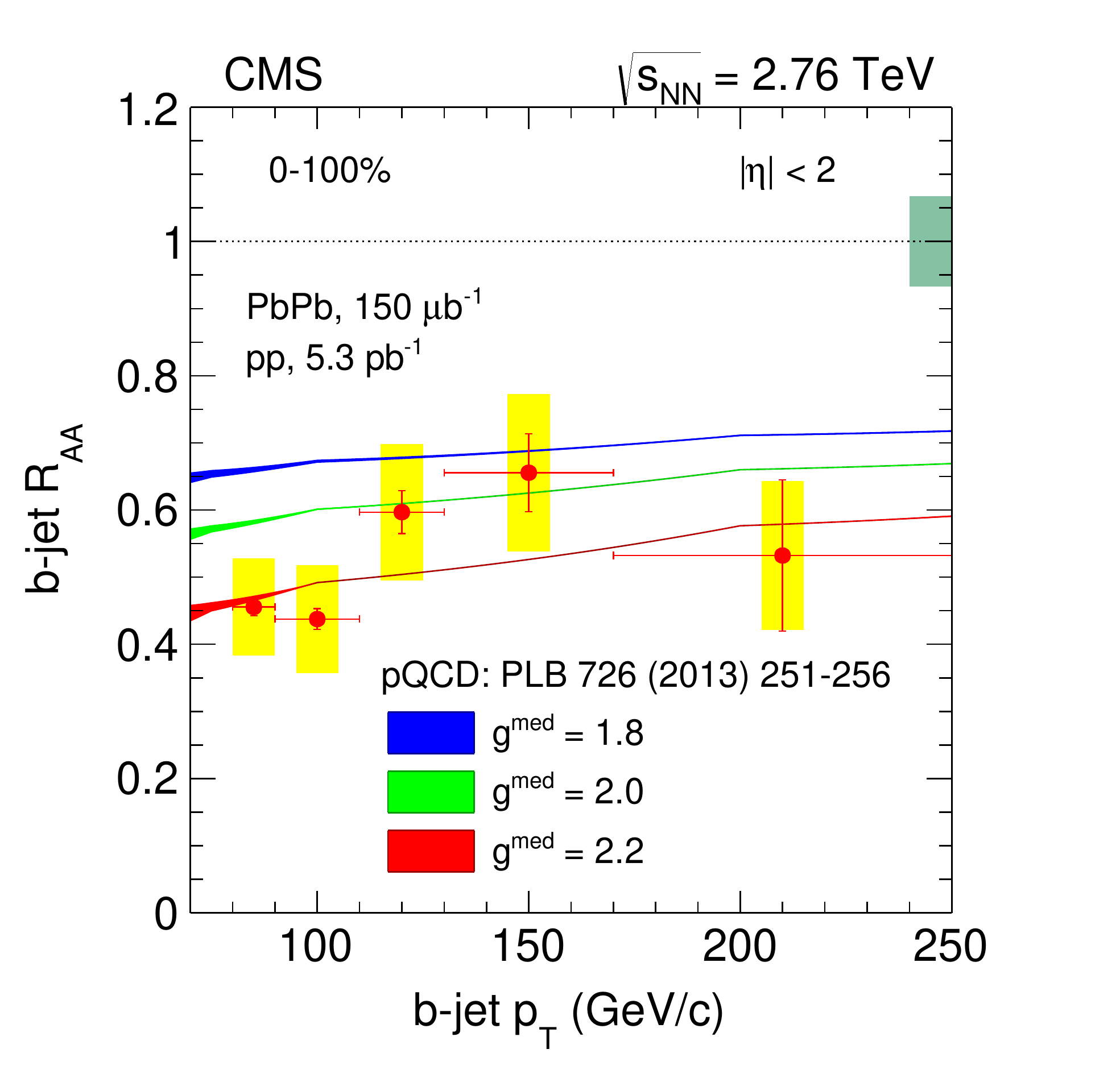}
\caption{\label{fig:raaVsPt}
The centrality integrated (0--100\%) \cPqb-jet $\raa$ as a function of \pt. Vertical and horizontal bars represent statistical uncertainties and bin widths, respectively, while filled boxes represent systematics uncertainties. The normalization uncertainty from the integrated luminosity in $\Pp\Pp$ collisions and from \taa is represented by the green band around unity.  The data are compared to pQCD-based calculations from~\cite{Huang:2013vaa}.
}
\end{center}
\end{figure}

Figure~\ref{fig:raaVsPt} shows the centrality-integrated \cPqb-jet nuclear modification factor ($\raa$), which is the ratio of the  \taa-normalized PbPb yield and the measured $\Pp\Pp$ cross section in Fig.~\ref{fig:spectra}, as a function of \pt. The jet and  \cPqb-tagging systematic uncertainties in $\raa$ are obtained by varying the $\Pp\Pp$ and PbPb data simultaneously.  This results in partial cancellation, giving a systematic uncertainty of 16--21\%, which is dominated by the  \cPqb-tagging uncertainty. A significant suppression of the yield with respect to the $\Pp\Pp$ expectation is observed in $\cPqb$ jets, which is indicative of the parton energy loss in the hot medium. No strong trend is observed as a function of \pt, although the data hint a modest rise at higher \pt. The data are compared to pQCD-based calculations from~\cite{Huang:2013vaa}.  The data are found to be consistent with a jet-medium coupling ($g^{med}$) in the range of 1.8--2, similar to the value found for inclusive jets.

\begin{figure}[htb]
\begin{center}
\includegraphics[width=0.48\textwidth]{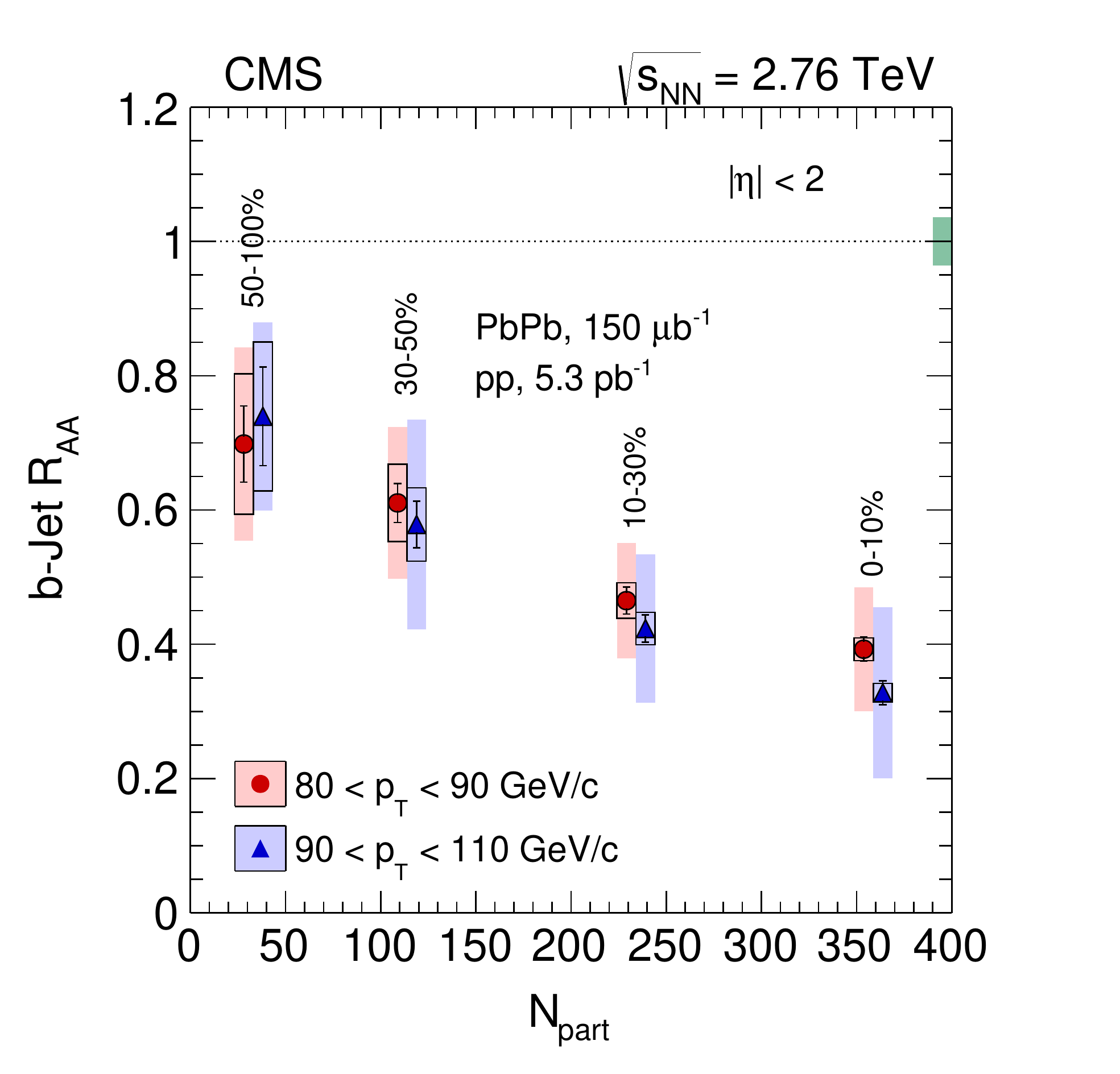}
\caption[\cPqb-jet $\raa$ vs. $N_\text{part}$]{\label{fig:raaVsCent}
The \cPqb-jet $\raa$, as a function of $N_\text{part}$ for two jet \pt\ selections as indicated in the legend.  Statistical uncertainties are shown as error bars. The filled boxes represent the systematics uncertainties, excluding the  \taa  uncertainties, which are depicted as open boxes. The normalization uncertainty in the integrated luminosity in $\Pp\Pp$ collisions is represented by the green band around unity.
}
\end{center}
\end{figure}

Figure~\ref{fig:raaVsCent} shows $\raa$ as a function of the number of participating nucleons ($N_\text{part}$), which is derived from the centrality (as measured by the energy in the forward calorimeters) through a Glauber calculation. Data for $80 < \pt < 90$\GeVc and $90 < \pt < 110$\GeVc are shown.  For both jet selections $\raa$ shows a smooth decrease with increasing centrality from about 0.70--0.75 to about 0.35--0.40.

The data presented in this study demonstrate the jet quenching phenomenon in the \cPqb-jet sector using fully reconstructed $\cPqb$ jets for the first time in heavy-ion collisions. Integrating over all collision centralities, $\cPqb$ jets are found to be suppressed over the 80--250\GeVc \pt range explored in this study. For the 80--110\GeVc \pt range, $\raa$ is found to decrease with collision centrality.  At larger \pt, the trend is less evident due to the reduced statistical precision.  The \cPqb-jet suppression is found to be qualitatively consistent with that of inclusive jets~\cite{Aad:2012vca}.  Although a sizeable fraction of \cPqb-tagged jets come from gluon splitting, a large mass and/or flavor dependence for parton energy loss can be excluded.  For example, a model based on strong coupling (via the AdS/CFT correspondence)~\cite{Horowitz:2007su}, in which mass effects could persist to large \pt would be incompatible with the current data, in contrast to a perturbative model in which mass effects are expected to be small at large \pt~\cite{Huang:2013vaa}.  A milder mass dependence, but one which still persists to large \pt, as predicted for light and heavy flavor hadrons in Ref.~\cite{Buzzatti:2011vt}, cannot be ruled out with the present uncertainties.

We congratulate our colleagues in the CERN accelerator departments for the excellent performance of the LHC and thank the technical and administrative staffs at CERN and at other CMS institutes for their contributions to the success of the CMS effort. In addition, we gratefully acknowledge the computing centres and personnel of the Worldwide LHC Computing Grid for delivering so effectively the computing infrastructure essential to our analyses. Finally, we acknowledge the enduring support for the construction and operation of the LHC and the CMS detector provided by the following funding agencies: BMWF and FWF (Austria); FNRS and FWO (Belgium); CNPq, CAPES, FAPERJ, and FAPESP (Brazil); MES (Bulgaria); CERN; CAS, MoST, and NSFC (China); COLCIENCIAS (Colombia); MSES and CSF (Croatia); RPF (Cyprus); MoER, SF0690030s09 and ERDF (Estonia); Academy of Finland, MEC, and HIP (Finland); CEA and CNRS/IN2P3 (France); BMBF, DFG, and HGF (Germany); GSRT (Greece); OTKA and NIH (Hungary); DAE and DST (India); IPM (Iran); SFI (Ireland); INFN (Italy); NRF and WCU (Republic of Korea); LAS (Lithuania); CINVESTAV, CONACYT, SEP, and UASLP-FAI (Mexico); MBIE (New Zealand); PAEC (Pakistan); MSHE and NSC (Poland); FCT (Portugal); JINR (Dubna); MON, RosAtom, RAS and RFBR (Russia); MESTD (Serbia); SEIDI and CPAN (Spain); Swiss Funding Agencies (Switzerland); NSC (Taipei); ThEPCenter, IPST, STAR and NSTDA (Thailand); TUBITAK and TAEK (Turkey); NASU (Ukraine); STFC (United Kingdom); DOE and NSF (USA).

\bibliography{auto_generated}   

\cleardoublepage \appendix\section{The CMS Collaboration \label{app:collab}}\begin{sloppypar}\hyphenpenalty=5000\widowpenalty=500\clubpenalty=5000\textbf{Yerevan Physics Institute,  Yerevan,  Armenia}\\*[0pt]
S.~Chatrchyan, V.~Khachatryan, A.M.~Sirunyan, A.~Tumasyan
\vskip\cmsinstskip
\textbf{Institut f\"{u}r Hochenergiephysik der OeAW,  Wien,  Austria}\\*[0pt]
W.~Adam, T.~Bergauer, M.~Dragicevic, J.~Er\"{o}, C.~Fabjan\cmsAuthorMark{1}, M.~Friedl, R.~Fr\"{u}hwirth\cmsAuthorMark{1}, V.M.~Ghete, C.~Hartl, N.~H\"{o}rmann, J.~Hrubec, M.~Jeitler\cmsAuthorMark{1}, W.~Kiesenhofer, V.~Kn\"{u}nz, M.~Krammer\cmsAuthorMark{1}, I.~Kr\"{a}tschmer, D.~Liko, I.~Mikulec, D.~Rabady\cmsAuthorMark{2}, B.~Rahbaran, H.~Rohringer, R.~Sch\"{o}fbeck, J.~Strauss, A.~Taurok, W.~Treberer-Treberspurg, W.~Waltenberger, C.-E.~Wulz\cmsAuthorMark{1}
% ...
\vskip\cmsinstskip
% ...
1:~~Also at Vienna University of Technology, Vienna, Austria\\
2:~~Also at CERN, European Organization for Nuclear Research, Geneva, Switzerland\\
% ...

\end{sloppypar}
\end{document}